\documentclass[aps,prl,,superscriptaddress,twocolumn,showpacs]{revtex4-1}
\usepackage{amsmath, amssymb}
\usepackage{color}
\usepackage{times}
\usepackage{graphicx}

\newtheorem{observation}{Observation}
\newtheorem{proposition}{Proposition}
\newtheorem{corollary}{Corollary}

\newtheorem{theorem}{Theorem}
\newtheorem{lemma}{Lemma}

\newcommand{\proof}{{\bf Proof. }}

\begin{document}

\title{Bipartite bound entanglement in continuous variables through deGaussification}

\author{F. E. S. Steinhoff} \email{steinhof@ifi.unicamp.br}
\affiliation{Arbeitsgruppe Quantenoptik, Institut f\"ur Physik, Universit\"at Rostock, D-18051 Rostock, Germany}
\affiliation{Instituto de F\'\i
sica ``Gleb Wataghin'', Universidade Estadual de Campinas,
13083-970, Campinas, SP, Brazil}
\author{M. C. de Oliveira }\email{marcos@ifi.unicamp.br}
\affiliation{Instituto de F\'\i
sica ``Gleb Wataghin'', Universidade Estadual de Campinas,
13083-970, Campinas, SP, Brazil}
\affiliation{Institute for Quantum Information Science, University of Calgary, Alberta T2N 1N4, Canada}
\author{J. Sperling} \email{jan.sperling2@uni-rostock.de}
\affiliation{Arbeitsgruppe Quantenoptik, Institut f\"ur Physik, Universit\"at Rostock, D-18051 Rostock, Germany}
\author{W. Vogel}
\affiliation{Arbeitsgruppe Quantenoptik, Institut f\"ur Physik, Universit\"at Rostock, D-18051 Rostock, Germany}

\date{\today}

\begin{abstract}
We introduce a class of  bipartite entangled continuous variable states that are positive under partial transposition operation, i.e., PPT bound entangled. These states are based on realistic preparation procedures in optical systems, being thus a feasible option to generate and observe genuinely bipartite bound entanglement in high precision experiments. One fundamental step in our scheme is to perform a non-Gaussian operation over a single-mode Gaussian state; this deGaussification procedure is achieved through a modified single-photon addition, which is a procedure that has currently being investigated in diverse optical setups. Although dependent on a single-photon detection in a idler channel, the preparation can be made unconditional after a calibration of the apparatus. The detection and proof of bound entanglement is made by means of the Range Criterion, theory of Hankel operators and Gerschgorin Disk's perturbation theorems. 
\end{abstract}
\maketitle

The characterization of the phenomenon known as bound entanglement \cite{horodecki1} constitutes one of the greatest challenges in quantum information theory \cite{book2}. Bound entangled states are those that, although being entangled, do not allow distillation of any pure entangled state with Local Operations and Classical Communication (LOCC). This kind of entanglement is associated with non-intuitive theoretical aspects of quantum information processing, such as communication using zero capacity channels \cite{smith}, or the irreversibility of entanglement under LOCC \cite{yang, brandao,mfc}. They also have practical implication in quantum cryptography \cite{horodecki2}, channel discrimination \cite{piani} and many quantum information protocols in general \cite{masanes}. Experimental realization of such states has been only recently achieved \cite{amselem,lavoie,barreiro,eisert} and is restricted until now to the multipartite scenario. Particularly, for continuous variable systems, local Gaussian operations, which are relatively simple to implement experimentally, are not able to distill entanglement if the system state is Gaussian \cite{nondistill,nondistill1}. In fact it is impossible to generate bound entanglement with bimodal Gaussian states \cite{ww}. Quite recently an experimental investigation explored this fact for the generation of such a kind of states employing a bipartition with more than two modes at Alice and Bob sides \cite{eisert}. To generate a two-mode continuous variable bound entangled state  one necessarily has to move it out from the Gaussian class of states, by implementing some non-Gaussian operation over a Gaussian state (\textit{deGaussification}).

In this article we propose a class of genuinely (two-mode) bipartite bound entangled states of arbitrary dimension that can be unconditionally prepared in optical systems with simple extension on current experimental techniques. Our approach is to \textit{deGaussify} a thermal state by a photon-addition followed by an incoherent mixing with a squeezed vacuum in an orthogonal polarization. These states serve as the inputs for generation of entanglement when mixed with vacuum states. The detection of entanglement is simple and is made using the well-known Range Criterion \cite{range}. Choosing properly the parameters involved, it is possible to design a state that is Positive under Partial Transposition (PPT), hence undistillable \cite{horodecki3}, representing a practical improvement over the findings of Ref. \cite{hlewenstein}. As a side result  we derive some theoretical insights on one-mode states through connections between PPT property and Hankel operator theory (see Proposition 1), which are used together with Hadamard products and Gerschgorin Disk's perturbation theorems for the proof of bound entangled states.

\emph{Preparation.} It is impossible to generate a bound entangled state for a system composed of a
single mode for Alice and an arbitrary number of modes for Bob, if their joint state $\rho_{AB}$ is Gaussian - Positivity of the partial transpose of $\rho_{AB}$ sufficiently implies
separability \cite{ww}. The sufficiency no longer holds if Alice has more than one mode, such as in the experimental implementation on Ref. \cite{eisert}, or if the joint state is not Gaussian. The procedure employed here for generating bound entanglement between two modes employs the later approach, and requires two fundamental steps - a photo-addition over a thermal state, and an incoherent mixture to a two-mode squeezed vacuum. The experimental setup is presented in Fig. \ref{fig1}.
\begin{figure}[htbp]
\begin{center}
\includegraphics[width=.45\textwidth]{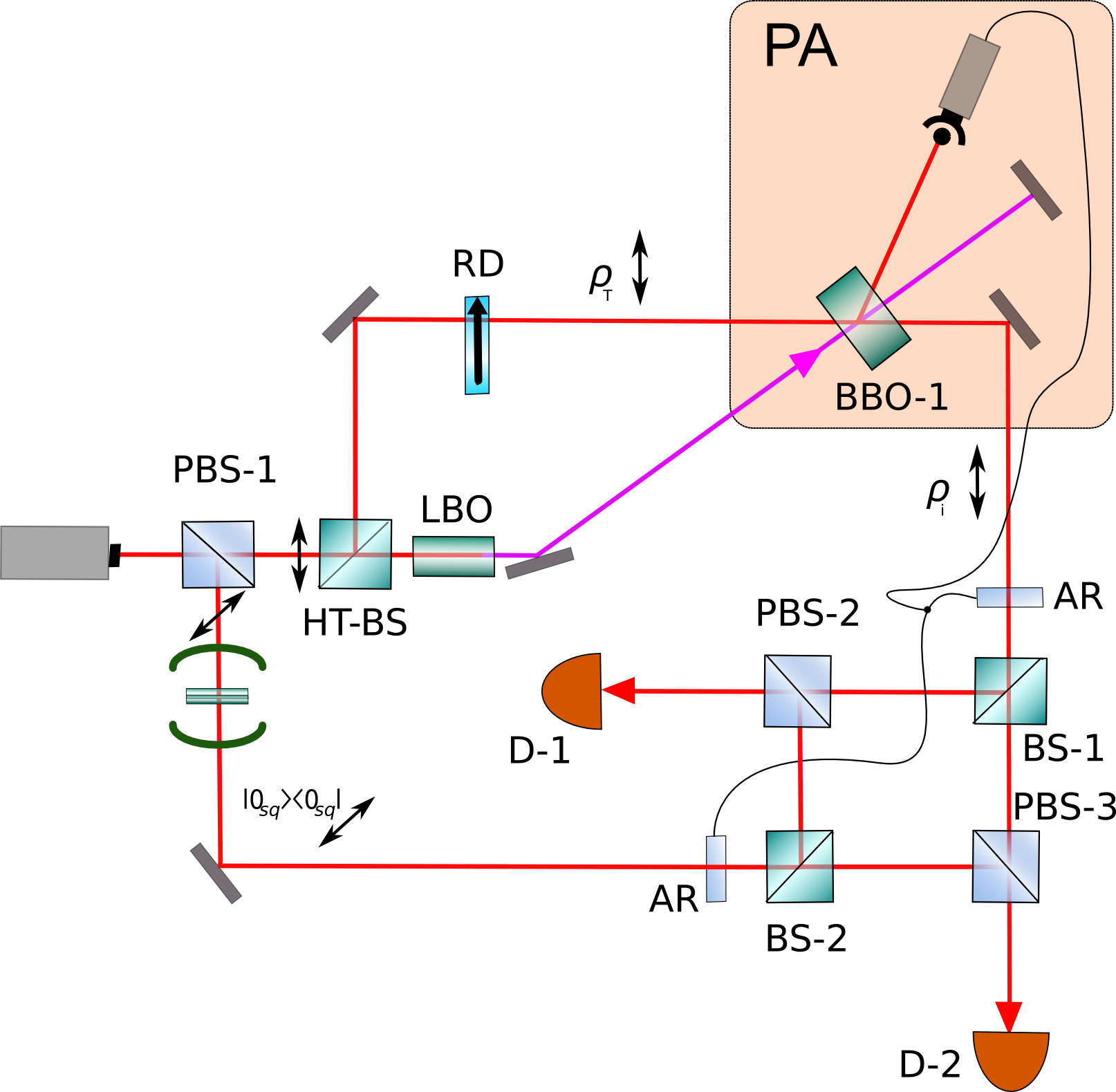} {}\label{fig1}
\end{center}
\caption{(Color Online) Two orthogonally polarized components of a laser beam are split in the polarizing beam splitter PBS-1. The $V$-polarized beam is prepared in a thermal state $\rho_T$ through the rotating disc RD and photon-added in the process described in the box PA (see text for details) to originate state $\rho_i$. This mode is sent through the 50:50 beam splitter BS-1 whose output is described by Eq. (\ref{out1}). The $H$-polarized component of the laser beam coming out of PBS-1 is used through an optical parametric oscillator to generate a squeezed vacuum state $|0_{sq}\rangle$, which is sent through BS-2 whose output is given by Eq. (\ref{squeezed}). The $V$-polarized and $H$-polarized output modes of BS-1 and BS-2, respectively, are sent through PBS-2 and PBS-3 to homodyne detection processes D-1 and D-2. The active retarders (or modulators) AR allow that only one of the $V$ or $H$ polarizations to proceed to D-1 and D-2 conditioned to photon-addition process in PA and on external control. The resulting state coming out from reconstructions in D-1 and D-2 is given by Eq. (\ref{gclass}).}\label{fig1}
\end{figure}
There a $V$-polarized single-mode is prepared in a thermal state, $\rho_T = \sum_n p_n|n\rangle\langle n|$, with thermal distribution $\{p_n\}$, through a phase-randomization process by a rotating disk (RD)\cite{bellini}. This light mode, $A$, is then photon-added (through a process inside the box PA to be later described), transforming $\{p_n\}$ into $\{p'_n\}$, and then mixed with a vacuum state mode, $B$, on the $50:50$ beam-splitter BS-1. Since an arbitrary beam-splitter action on the creation operators of two input modes is given by the global unitary operation 
\begin{eqnarray} 
U(\theta)\left(\begin{array}{c}
{{a_A}^\dagger} \\ 
{{a_B}^\dagger}\end{array}\right)=
\left(\begin{array}{c c}
{cos(\theta)}&{sin(\theta)} \\
{-sin(\theta)}&{cos(\theta)}\end{array}\right)\left(\begin{array}{c}
{{a_A}^\dagger} \\ 
{{a_B}^\dagger}\end{array}\right) \label{bs},
\end{eqnarray} 
the output state for the $50:50$  ($\theta = \pi/4$)  beam-splitter BS-1 is given by
\begin{eqnarray}
\rho = U\left(\frac{\pi}{4}\right)\left(\rho_i\otimes |0\rangle\langle 0|\right)U^{\dagger}\left(\frac{\pi}{4}\right) = \sum_{n=0}^{\infty}p'_n|\psi_{n,0}\rangle\langle\psi_{n,0}|,\label{out1}
\end{eqnarray}
where 
\begin{equation}|\psi_{n,0}\rangle=U(\pi /4)|n,0\rangle=\frac{1}{2^{n/2}}\sum_{k=0}^ n {n\choose k}^{1/2}|k,n-k\rangle,\label{eqpsin}\end{equation}
are the eigenvectors of the density matrix $\rho$.  $\{|\psi_{n,0}\rangle\}$ - with $n=0,1,\ldots,\infty$ - form an orthonormal set, since the unitary $U$ preserves inner product, and are also permutationally invariant, i.e., given $\Pi=\sum_{i,j}|i,j\rangle\langle j,i|$,  $\Pi|\psi_{n,0}\rangle=|\psi_{n,0}\rangle$. This implies that $\Pi\rho=\rho \Pi=\rho$. The state (\ref{out1}) alone cannot generate PPT-bound entangled states, as shown in reference \cite{simon}. We thus consider a more general class of states, given by
\begin{eqnarray}
\rho'=\lambda\rho +(1-\lambda)|\Omega\rangle\langle\Omega|, \label{gclass}
\end{eqnarray}
corresponding to the mixture of state $\rho$ in (\ref{out1}) with the state 
\begin{eqnarray}
|\Omega\rangle = (1-|\omega|^2)^{1/4}\sum_{k=0}^{\infty}\sqrt{\frac{\Gamma(n+1/2)}{n!\sqrt{\pi}}}{\omega}^k|\phi_{2k,0}\rangle \label{squeezed}, 
\end{eqnarray}
whose preparation is depicted in Fig. 1 as we now explain. State (\ref{squeezed}) is generated by passing a $H$-polarized one-mode squeezed vacuum state,
\begin{eqnarray}
|0_{sq}\rangle = (1-|\omega|^2)^{1/4}\sum_{k=0}^{\infty}\sqrt{\frac{\Gamma(n+1/2)}{n!\sqrt{\pi}}}{\omega}^k|2k\rangle,
\end{eqnarray}
generated in an optical parametric oscillator, 
through the beam-splitter BS-2. Here $\omega=(\xi/|\xi|)\tanh(|\xi|/2)$ is the squeezing parameter and thus $0<|\omega| <1$.
 We impose for BS-2 that the parameter $\theta$ in (\ref{bs}) satisfy $\theta\neq\pi/4$, i.e, not a $50:50$ beam-splitter. For simplicity, we take $0<\theta<\pi/4$. Thus we have
$|\Omega\rangle=U(\theta)|0_{sq}\rangle$,
and writing $|\phi_{n,0}\rangle = U(\theta)|n,0\rangle$, we arrive at expression (\ref{squeezed}), with 
\begin{equation}|\phi_{2k,0}\rangle=\sum_{l=0}^{2k} {2k\choose l}^{1/2}(\cos \theta)^l (\sin\theta)^{2k-l} |l,2k-l\rangle.\end{equation}

Now to prepare the mixture in (\ref{gclass}), the two independent preparations are recombined in the polarizing beam-splitters PBS-2 and PBS-3 and proceed for homodyne detection on D-1 and D-2. The active retarders (AR) are externally controlled by the detection of a photon in the photo-addition process plus a external control to generate the full range of $\lambda$ in (\ref{gclass}). Whenever this photon is detected in the idler mode generated in the parametric dow-conversion in BBO-1, indicating that a photon has been added to the signal mode, the AR allow only the $V$-polarized photon-added thermal component to leave to the detectors. When no photon is detected only the $H$-polarized two-mode squeezed vacuum is allowed to proceed to the detectors. By neglecting some of those detections, the ARs control the fraction $\lambda$ of polarization of the field incident at the photodetectors for repeated experiments. The idea behind this preparation is to use the fact that the beam-splitter is a classicality-preserving device and converts a (non-)classical state into a (in)separable one. The mixing in (\ref{gclass}) is thus a mixture of two nonclassical states and since the mixing parameter $\lambda$ is controllable by the experimentalist through the ARs, it is possible to prepare an entangled state whose partial transposition is still positive. As we will see a crucial element here was the elimination of the vacuum component from the thermal state by the photon-addition process.

\textit{Photon added and shifted thermal states.} For the mixture present in Eq. (\ref{gclass}) to allow a genuinely bound entangled state it is necessary to first change the Gaussian character of the $V$-polarized thermal state at mode $A$. A simple procedure to deGaussify the thermal state is the photon-addition as described in \cite{bellini}, which assumes that when the thermal state is fed as a signal into a parametric amplifier, the output signal state is conditionally prepared every time that a single photon is detected in the correlated idler mode. The simple assumption here is that the action of the conditioned parametric amplification is given up to first order in the coupling $g$ between idler and signal by $ [1+(g a^\dagger_s a^\dagger_i-g^* a_s a_i)]$ where $a^\dagger_{s(i)}$ are bosonic creation operators acting on the signal (idler) modes. This results in a photon-added thermal state, which although possessing the character needed (absence of the vacuum state), has failed to
produce a PPT state. However if instead one is able to implement a saturated photon-addition  \cite{photocount} in the sense that the action of the creation and annihilation operators in the signal is replaced by $E^+=a_s^\dagger(a_s^\dagger a_s+1)^{-1/2}$ and $E^-=(a_s^\dagger a_s+1)^{-1/2}a_s$ respectively, the resulting state conditioned to the detection of one photon in the idler is given by the shifted-thermal state  \cite{lee}, 
\begin{eqnarray}
\rho_i=\sum_{n=0}^{\infty}\frac{1}{\bar{n}+1}\left(\frac{\bar{n}}{\bar{n}+1}\right)^n|n+1\rangle\langle n+1|, \label{input1}
\end{eqnarray}
where $\bar{n}$ is the mean thermal photon number. These states were first considered by Lee in reference \cite{lee}, in an analysis of the scheme depicted in \cite{scully}, where laser cooling significantly changes the emission of radiation of a micromaser. The distribution (\ref{input1}) arises in \cite{lee} when the parameters of the micromaser cavity fulfill an ideal requirement. The shifting operation is also present in the related proposal of \cite{vogel}. Remarkably, a scheme to prepare (\ref{input1}) deterministically has been recently proposed \cite{oi}, making the unconditional preparation of $(\ref{gclass})$ an achievable goal. The shifted thermal state allows the state (\ref{gclass}) to have genuine bound entanglement as we discuss in what follows.

\emph{Entanglement detection.} We now give a set of general rules for determining an arbitrary PPT bound entangled state.
\begin{observation}
State (\ref{gclass}) is entangled for any $0<|\omega|<1$.
\end{observation}
\textbf{Proof:} 
We use the following theorem, known as the Range Criterion \cite{range}:
\begin{theorem}
For every separable state, there exists a set of product vectors $\{|a,b\rangle\}$ which spans the range of its density matrix, such that $\{|a,b^*\rangle\}$ spans its partial transposition. 
\end{theorem}
A violation of at least one of the conditions of this criterion implies entanglement, thus we prove that the range of $\rho'$ in (\ref{gclass}) does not contain a single product vector. 
The range of $\rho'$ is spanned by $\{|\psi_{n,0}\rangle\}$ - with $n=0,1,\ldots,\infty$ - and $|\Omega\rangle$ as given in Eqs. (\ref{eqpsin}) and (\ref{squeezed}), respectively. We show now that assuming a product vector in the range of $\rho'$ leads to a contradiction. 

Thus, let us assume that there exist complex numbers $m_k$, $k=0, 1, 2, \ldots$ such that
\begin{eqnarray}
m_0|\Omega\rangle +\sum_{k=1}^{\infty}m_k|\psi_{k,0}\rangle= |v_1\rangle\otimes |v_2\rangle \label{assumption}
\end{eqnarray}
where 
$|v_1\rangle = \sum_{k=0}^{\infty}\alpha_k|k\rangle$,
$|v_2\rangle = \sum_{k'=0}^{\infty}\beta_{k'}|k'\rangle$. 
Let us write explicitly some important terms in left-hand side of (\ref{assumption}):
\begin{eqnarray*}
& & m_0|0,0\rangle + m_1\left(|0,1\rangle+|1,0\rangle\right) \\
&+& \left((\omega/\sqrt{2})m_0\cos^2\theta+m_2\right)|0,2\rangle \\
&+& \left((\omega/\sqrt{2})m_0\sin\theta\cos\theta+\sqrt{2}m_2\right)|1,1\rangle \\
&+&\left((\omega/\sqrt{2})m_0\sin^2\theta+m_2\right)|2,0\rangle\\
&+& m_3\left(|0,3\rangle +\sqrt{3}|1,2\rangle+\sqrt{3}|2,1\rangle+|3,0\rangle\right) \\
&+& \left(m_0\omega^2(3/8)\cos\theta\sin^3\theta +m_4\sqrt{4} \right)|1,3\rangle \\
&+& \left(m_0\omega^2(3/8)\cos^3\theta\sin\theta +m_4\sqrt{4} \right)|3,1\rangle \\
&+& \ldots
\end{eqnarray*}
We disregard normalization factors, since one can always incorporate these factors in the values $m_i$. The first terms in the right-hand side of (\ref{assumption}) are
\begin{eqnarray*}
& & \alpha_0\beta_0|0,0\rangle +\alpha_0\beta_1|0,1\rangle + \alpha_1\beta_0|1,0\rangle \\
&+& \alpha_0\beta_2|0,2\rangle +\alpha_1\beta_1|1,1\rangle + \alpha_2\beta_0|2,0\rangle \\
&+& \alpha_0\beta_3|0,3\rangle +\alpha_1\beta_2|1,2\rangle +\alpha_2\beta_1|2,1\rangle +\alpha_3\beta_0|3,0\rangle \\
&+& \alpha_1\beta_3|1,3\rangle + \alpha_3\beta_1|3,1\rangle\\
&+& \ldots
\end{eqnarray*}
Let us first assume $m_0=0$, this implies $\alpha_0\beta_0=0$, by the linear independence of the set $\{|i,j\rangle\}$ - with $i,j=0,1,\ldots,\infty$. Hence either $\alpha_0=0$ or $\beta_0=0$; it is straightforward that either case would imply $m_i=0$ for all $i$, clearly being a contradiction. Let us then consider the case $m_0\neq 0$. Without loss of generality, we assume $\alpha_0=\beta_0=1$; from expression (\ref{assumption}), one gets $m_0=1$ and since $\{|\psi_{n,0}\rangle\}$ is permutationally invariant, for all $n=0,1,\ldots,\infty$, we have for the first odd terms 
$m_1=\alpha_1=\beta_1$, and $m_3=\alpha_3=\beta_3$. 
So $\alpha_1\beta_3=\alpha_3\beta_1$. However, we have for the corresponding terms
$\alpha_1\beta_3 = m_0\omega^2(3/8)\cos\theta\sin^3\theta +m_4\sqrt{4}$, and
$\alpha_3\beta_1 = m_0\omega^2(3/8)\cos^3\theta\sin\theta +m_4\sqrt{4}$.
Since $\theta\neq\pi/4$, we conclude that $\alpha_3\beta_1\neq\alpha_1\beta_3$, a contradiction \cite{nota3}. Thus, there is no product vector in the range of $\rho'$, implying Observation 1 is true, QED.

As pointed out in \cite{hlewenstein}, it is possible to construct entanglement witnesses for these states by applying the optimization methods of references \cite{sperling} to projectors over the range of $\rho'$. Let us now check the conditions to obtain a PPT state. The following proposition relates the photocounting distribution of the input state $\rho_i$ to the PPT property of the output state (\ref{out1}).
\begin{proposition}
State (\ref{out1}) is PPT if the following infinite-dimensional Hankel matrix \cite{nota1}.
\begin{eqnarray}
A_0 = \left(\begin{array}{c c c c}
{p_0}&{p_1}&{p_2}&{\ldots} \\
{p_1}&{p_{2}}&{p_3}&{\ldots} \\
{p_2}&{p_3}&{p_4}&{\ldots} \\
{\vdots}&{\vdots}&{\vdots}&{\ddots}
\end{array}\right),
\end{eqnarray}
is positive semidefinite.
\end{proposition}
Details of the proof are left for the Appendices A and B. A crucial point is that under a special ordering of the basis of the total Hilbert space $\mathcal{H}_A\otimes\mathcal{H}_B$, the partially transposed matrix of (\ref{out1}) assumes a block-diagonal structure, 
\begin{eqnarray}
\rho^{T_B} = \bigoplus_{i=-\infty}^{\infty}M_i \label{structure1},
\end{eqnarray}
where each block can be decomposed as a Hadamard product of two matrices $M_i=A_i\circ B_i$, being both
\begin{eqnarray}
A_i = \left(\begin{array}{c c c c}
{p_{|i|}}&{p_{|i|+1}}&{p_{|i|+2}}&{\ldots} \\
{p_{|i|+1}}&{p_{|i|+2}}&{p_{|i|+3}}&{\ldots} \\
{p_{|i|+2}}&{p_{|i|+3}}&{p_{|i|+4}}&{\ldots} \\
{\vdots}&{\vdots}&{\vdots}&{\ddots}
\end{array}\right),
\end{eqnarray}
and $B_i$ positive matrices described in Appendix B. Since the Hadamard product preserves positivity and $B_i$ is positive, $\rho^{T_B}$ will be positive semidefinite if $A_i$ is positive semidefinite. But through Sylvester's criterion \cite{horn} one sees that this condition can be reduced to positive semidefiniteness of $A_0$. Thus the positivity under partial transposition can be checked by a hierarchical sequence of photocounting probabilities matrices. Indeed, a similar connection is already present in reference \cite{simon} with regard to the Stieltjes moment problem, in terms of nonclassicality exhibited by states Negative under Partial Transposition (NPT). An advantage of our approach is the direct-sum structure in Eq. (\ref{structure}), which allows us to handle the construction of PPT states in a simple way. Similar direct-sum decompositions of the partially transposed density matrix can be found in \cite{steinhoff, chruscinski} and bring a great deal of simplification when dealing with PPT bound entanglement. 

The idea to obtain a PPT (\ref{gclass}) is that the only block (\ref{structure1}) for the shifted-thermal state that is negative under partial transposition is the block $M_0$, due to $p_0=0$. Hence, with a suitable small squeezing parameter $\omega$, the vacuum amplitude of the squeezed state, i.e $|\langle 00|\Omega_{sq}\rangle|^2=\sqrt{1-|\omega|^2}$ will be suitably big in such a way to replace the null vacuum amplitude of $\rho$, making the mixture $\rho'$ positive under partial transposition. The other terms are $O(|\omega|)$, being as small as one desires, representing a small perturbation under control. We obtain thus:
\begin{observation}
There exists states (\ref{gclass}) which are PPT. By Observation 1 these states are bound entangled. 
\end{observation}
It is intuitive that Observation 2 is true given that the eigenvalues of a matrix are continuous functions of its entries and thus a small perturbation of these values will not change the eigenvalues significantly. A constructive proof of Observation 2 is left for the Appendix C, where we consider the example of such states with a balanced mixture $\lambda = 1/2$ and a shifted-thermal state $\bar{n}=1$, that is
\begin{eqnarray}
\rho' = \frac{1}{2}\left(\rho+|\Omega\rangle\langle\Omega|\right), 
\end{eqnarray}
with
\begin{eqnarray}
\rho_i=\sum_{n=1}^{\infty}\left(\frac{1}{2}\right)^n|n\rangle\langle n|.
\end{eqnarray}
Although in practice, for continuous variable's regime it is analytically and numerically hard to determine exactly \cite{nota4} the spectrum of $\rho'$, it is possible to obtain upper bounds for the values of $|\omega|$, which guarantee a positive partial transposition. For the example considered, a conservative upper bound of $|\omega|\leq 10^{-3}$ has proved to be sufficient, allowing the generation of a bound entangled state in continuous variables. It will remain an open question whether we have found or not a \textit{generic} continuous variable bipartite bound entangled state, i.e., a bound entangled state with infinite Schmidt rank \cite{hlewenstein}. 
 
We have presented a procedure to unconditionally prepare bipartite bound entangled states in continuous variable's regime. The approach assumed was to deGaussify a thermal field by a photon addition process and to mix it with a squeezed vacuum state. Several minor results were developed in order to achieve this goal. Particularly the links with Hankel operator theory and the direct-sum structure of the partially transposed state allowed us to give bounds on the parameters that enable PPT bound entangled states to be produced. There are few examples of such states in continuous variables and thus the novel class (\ref{gclass}) is interesting in its own, even if it were impossible to envisage a scheme to generate it experimentally. Our work shows that this is not the case and that the preparation in practice is feasible, opening new possibilities in quantum information processing protocols, as well as in the theory of quantum entanglement. We focused on thermal fields, due to their implementation simplicity in the laboratory and also due to their use in fundamental experiments as \cite{bellini}, but we stress that similar results could in principle be achieved with different classical photocounting distributions. Our approach was to "kill" the vaccum contribution of the thermal state by performing a photo-addition and then replacing it with the vaccum term of the squeezed state, which is superposed with other terms. If one is able to produce a one-mode state satisfying Proposition 1 and then could entangle its vaccum term with other convenient state, we believe it is possible to construct similar states to (\ref{gclass}); this point is currently being investigated.  

\section*{Acknowledgment}
This work was supported by the Deutsche Forschungsgemeinschaft through SFB 652 and the brazilian agency CAPES, through PDEE program, Process $2021-10-2$. MCO acknowledges support from AITF and the Brazilian agencies CNPq and FAPESP
	through the Instituto Nacional de Ci{\^e}ncia e Tecnologia --- Informa\c{c}{\~a}o Qu{\^a}ntica (INCT-IQ). FESS is specially grateful for the hospitality and stimulating discussions of Prof. Vogel's group during his visit to the University of Rostock. \\

\appendix
\section{Appendix A: Block structure of output}

In terms of Fock basis, the state (\ref{out1}) reads
\begin{eqnarray}
\rho = \sum_{n=0}^{\infty} p_n\sum_{k,k'=0}^n c^{(n)}_{k,k'}|k,n-k\rangle\langle k',n-k'|,
\end{eqnarray}
where we introduced the symbols $c^{(n)}_{k,k'}= \frac{1}{2^k}\sqrt{{n\choose k}{n\choose k'}}$. We proceed now in order to find the structure of the partial transposition of the density matrix $\rho$, which is obtained performing the operation of transposition in only one of the subsystems. From expression (\ref{out1}) and using permutation invariance, 
\begin{eqnarray}
\rho = \sum_{n=0}^{\infty} p_n\sum_{k,k'=0}^n c^{(n)}_{k,k'}|k,n-k\rangle\langle n-k',k'|,
\end{eqnarray}
hence the partial transposed matrix of the second mode in Fock basis is given by
\begin{eqnarray}
\rho^{T_B}= \sum_{n=0}^{\infty}p_n\sum_{k,k'=0}^n c^{(n)}_{k,k'}|k,k'\rangle\langle n-k',n-k|. \label{ppt1}
\end{eqnarray}
An arbitrary element of this matrix is
\begin{eqnarray*}
\langle a,b|\rho^{T_B} |c,d\rangle = \sum_n p_n\sum_{k,k'=0}^n c^{(n)}_{k,k'}\delta_{a,k}\delta_{b,k'}\delta_{c,n-k'}\delta_{d,n-k},
\end{eqnarray*}
and we have then $\langle a,b|\rho^{T_B} |c,d\rangle = 0$ unless $a+d=c+b=n$, or, equivalently, $a-b=c-d$. Taking this rule into account, we choose a special ordering of the basis of the Hilbert space $\mathcal{H}=\mathcal{H}_A\otimes\mathcal{H}_B$ so that the matrix $\rho^{T_B}$ has a special block structure in this ordering. Let us first define the following sets:
\begin{eqnarray} 
\mathcal{B}_0 &=& \{|jj\rangle\}_{j=0}^{\infty} , \\
\mathcal{B}_{+i} &=& \{|i+k,k\rangle\}_{k=1}^{\infty}, \\
\mathcal{B}_{-i} &=& \{|k,i+k\rangle\}_{k=1}^{\infty}.
\end{eqnarray}
The notation should be clear: the elements in the set $\mathcal{B}_0$ are vectors $|ab\rangle$ which fulfill $a-b=0$, while elements in set $\mathcal{B}_{\pm i}$ are those which respect $a-b = \pm i$. The union of these sets is precisely the basis of the total Hilbert space $\mathcal{H}$, but now in a different ordering, i.e., we are ordering vectors $|ab\rangle$ according to their difference $a-b$. If we take as our ordered basis $\mathcal{B}=\bigcup_i\mathcal{B}_i$, the partially transposed matrix will show the block structure 
\begin{eqnarray}
\rho^{T_B} = \left(\begin{array}{c | c | c | c | c}
{\ddots}&{}&{}&{}&{} \\ \hline
{}&{M_{-1}}&{}&{}&{} \\  \hline
{}&{}&{M_{0}}&{}&{} \\  \hline
{}&{}&{}&{M_{+1}}&{} \\  \hline
{}&{}&{}&{}&{\ddots} 
\end{array}\right),
\end{eqnarray}
and we write
\begin{eqnarray}
\rho^{T_B} = \bigoplus_{i=-\infty}^{\infty}M_i,
\end{eqnarray}
with each block $M_i$ having the matricial representation
\begin{eqnarray*}
M_{\pm i}\stackrel{\mathcal{B}_{\pm i}}{=} \left(\begin{array}{c c c c}
{p_ic^{(i)}_{i,0}}&{p_{i+1}c^{(i+1)}_{i,0}}&{p_{i+2}c^{(i+2)}_{i,0}}&{\ldots} \\
{p_{i+1}c^{(i+1)}_{i+1,1}}&{p_{i+2}c^{(i+2)}_{i+1,1}}&{p_{i+3}c^{(i+3)}_{i+1,1}}&{\ldots} \\
{p_{i+2}c^{(i+2)}_{i+2,2}}&{p_{i+3}c^{(i+3)}_{i+2,2}}&{p_{i+4}c^{(i+4)}_{i+2,2}}&{\ldots} \\
{\vdots}&{\vdots}&{\vdots}&{\ddots} 
\end{array}\right),\label{form}
\end{eqnarray*}
where the notation $X\stackrel{\mathcal{C}}{=}$ means the matricial representation of $X$ in the basis $\mathcal{C}$. Thus, the state $\rho$ will be PPT iff all the blocks $M_i$ are positive semidefinite for all values $i\in Z$. 

To show the direct-sum decomposition in another way, we define $\mathcal{H}_i=Span\{\mathcal{B}_i\}$ ($Span\{.\}$ amounts to the linear span of $\{.\}$) and we have the direct-sum decomposition of the total Hilbert space as
\begin{eqnarray} 
\mathcal{H}=\bigoplus_{i=\infty}^{\infty}\mathcal{H}_i .
\end{eqnarray}
The subspaces $\mathcal{H}_i$ are invariant under the action of  the operator $\rho^{T_B}$, i.e., this operator does not send vectors from $\mathcal{H}_i$ to a different $\mathcal{H}_j$. So, the operator $\rho^{T_B}$ should decompose as a direct-sum\cite{hoffman}.

\section{Appendix B: Proof of Proposition 1} 

As shown in Appendix A, the partially transposed matrix of (\ref{out1}) assumes a block-diagonal structure, 
\begin{eqnarray}
\rho^{T_B} = \bigoplus_{i=-\infty}^{\infty}M_i \label{structure}.
\end{eqnarray}
Each block can be decomposed as a Hadamard product of two matrices $M_i=A_i\circ B_i$, with
\begin{eqnarray}
A_i = \left(\begin{array}{c c c c}
{p_{|i|}}&{p_{|i|+1}}&{p_{|i|+2}}&{\ldots} \\
{p_{|i|+1}}&{p_{|i|+2}}&{p_{|i|+3}}&{\ldots} \\
{p_{|i|+2}}&{p_{|i|+3}}&{p_{|i|+4}}&{\ldots} \\
{\vdots}&{\vdots}&{\vdots}&{\ddots}
\end{array}\right),
\end{eqnarray}
\begin{eqnarray} 
B_i = \left(\begin{array}{c c c c}
{c^{(i)}_{i,0}}&{c^{(i+1)}_{i,0}}&{c^{(i+2)}_{i,0}}&{\ldots} \\
{c^{(i+1)}_{i+1,1}}&{c^{(i+2)}_{i+1,1}}&{c^{(i+3)}_{i+1,1}}&{\ldots} \\
{c^{(i+2)}_{i+2,2}}&{c^{(i+3)}_{i+2,2}}&{c^{(i+4)}_{i+2,2}}&{\ldots} \\
{\vdots}&{\vdots}&{\vdots}&{\ddots} 
\end{array}\right).
\end{eqnarray}
We will need the following theorem, known as the \textit{Schur Product Theorem} \cite{horn}:
\begin{theorem}
The Hadamard product of two positive semidefinite matrices is a positive semidefinite matrix.
\end{theorem}
To prove Proposition 1, we prove first a lemma:
\begin{lemma}
The matrices $B_j$ are positive definite, for all $j\in Z$. 
\end{lemma}

\proof First, starting from the definition $c^{(k)}_{i,j}= \frac{1}{2^k}\sqrt{{k\choose i}{k\choose j}}=\frac{k!}{2^k}\sqrt{\frac{1}{i!j!(k-i)!(k-j)!}}$, we express an arbitrary $B_j$ as a Hadamard product of six matrices:
\begin{eqnarray}
B_j = C_j\circ D_j\circ E_j\circ E^{\dagger}_j\circ F_j\circ F^{\dagger}_j,
\end{eqnarray}
with
\begin{eqnarray*}
C_j &=& \left(\begin{array}{c c c c}
{j!}&{(j+1)!}&{(j+2)!}&{\ldots} \\
{(j+1)!}&{(j+2)!}&{(j+3)!}&{\ldots} \\
{(j+2)!}&{(j+3)!}&{(j+4)!}&{\ldots} \\
{\vdots}&{\vdots}&{\vdots}&{\ddots}
\end{array}\right) , \\
D_j &=& \left(\begin{array}{c c c c}
{1/2^j}&{1/2^{j+1}}&{1/2^{j+2}}&{\ldots} \\
{1/2^{j+1}}&{1/2^{j+2}}&{1/2^{j+3}}&{\ldots} \\
{1/2^{j+2}}&{1/2^{j+3}}&{1/2^{j+4}}&{\ldots} \\
{\vdots}&{\vdots}&{\vdots}&{\ddots}
\end{array}\right), \\
E_j &=&\left(\begin{array}{c c c c}
{1/\sqrt{0!}}&{1/\sqrt{0!}}&{1/\sqrt{0!}}&{\ldots} \\
{1/\sqrt{1!}}&{1/\sqrt{1!}}&{1/\sqrt{1!}}&{\ldots} \\
{1/\sqrt{2!}}&{1/\sqrt{2!}}&{1/\sqrt{2!}}&{\ldots} \\
{\vdots}&{\vdots}&{\vdots}&{\ddots}
\end{array}\right), \\
F_j &=&\left(\begin{array}{c c c c}
{1/\sqrt{j!}}&{1/\sqrt{j!}}&{1/\sqrt{j!}}&{\ldots} \\
{1/\sqrt{(j+1)!}}&{1/\sqrt{(j+1)!}}&{1/\sqrt{(j+1)!}}&{\ldots} \\
{1/\sqrt{(j+2)!}}&{1/\sqrt{(j+2)!}}&{1/\sqrt{(j+2)!}}&{\ldots} \\
{\vdots}&{\vdots}&{\vdots}&{\ddots}
\end{array}\right).
\end{eqnarray*}
The matrices $D_j$, $E_j$ and $F_j$ are rank-$1$ matrices, so are obviously positive. The Hankel matrix $C_0$ is positive definite, since its leading principal minors of order $k$ have determinant $\Pi_{i=0}^{k}(i!)^2$. A similar argument holds for an arbitrary $C_j$ and thus they are all positive definite. Another way to prove this is observing that the following sequence satisfies the Stieltjes moment problem \cite{shohat}:
\begin{eqnarray}f_n = n! = \int_0^{\infty}x^n e^{-x}dx, \ \ n=0,1,2,\ldots
\end{eqnarray}
So, $C_0$ is positive definite. Now, we see that
\begin{eqnarray}g_0&=&j!=\int_0^{\infty}x^0(x^j e^{-x})dx, \\g_1&=&(j+1)!=\int_0^{\infty}x^1(x^j e^{-x})dx, \\g_2&=&(j+2)!=\int_0^{\infty}x^2(x^j e^{-x})dx, \\\vdots \end{eqnarray}
defines a sequence that also satisfies the Stieltjes moment problem. From Theorem 1, the Hadamard product of positive-semidefinite matrices is a positive-semidefinite matrix. Also, the Hadamard product of a positive-definite matrix - $C_j$ in our case - with rank-$1$ matrices - $D_j$, $E_j$ and $F_j$ - is positive-definite \cite{nota2}. So $B_j$ is a positive-definite matrix, QED. 

The proof of Proposition 1 is now straightforward. Since all $B_i$ are positive definite, to have all $M_i$ positive semidefinite we must have all $A_i$ positive semidefinite. But $A_0$ positive semidefinite implies, by Sylvester's Criterion, that all other $A_i$ are positive semidefinite, since they are principal submatrices of $A_0$. So, all $M_i=A_i\circ B_i$ are positive semidefinite if $A_0$ is positive semidefinite and by the block structure of $\rho$ the state is PPT if $A_0$ is positive semidefinite, QED.

\section{Appendix C: Proof of Observation 2}

For simplicity, we will construct a example of a PPT (\ref{gclass}) with $\lambda=1/2$ and $\rho$ being the output of (\ref{input1}) with $\bar{n}=1$. Thus, the state we are considering is
\begin{eqnarray}
\rho' = \frac{1}{2}\left(\rho+|\Omega\rangle\langle\Omega|\right) \label{mix}
\end{eqnarray}
with
\begin{eqnarray}
\rho=\sum_{n=1}^{\infty}\left(\frac{1}{2}\right)^n|\psi_{n0}\rangle\langle \psi_{n0}| \label{exemplo}
\end{eqnarray}
We first observe that any shifted thermal state $\rho$ generated by (\ref{input1}) is locally equivalent to (\ref{exemplo}) above. Define the following invertible operation:
\begin{eqnarray} 
T|n\rangle = (\bar{n}+1)^{1/2}\left(\frac{\bar{n}+1}{2\bar{n}}\right)^{n/2}|n\rangle
\end{eqnarray}
Then we have that $T\otimes T\rho T^{\dagger}\otimes T^{\dagger}$ is equal to the matrix (\ref{exemplo}). Local invertible operations do not affect positivity of the partial transposed matrix, thus this special case is broad in this sense (for the case of $\lambda = 1/2$).

We know that (\ref{exemplo}) is NPT, since $p_0=0$, implying that $\rho_i$ is nonclassical and thus NPT, by the criterion of \cite{simon}. Also, from Appendix B, we know that
\begin{eqnarray}
\rho^{T_B} = \bigoplus_{i=-\infty}^{\infty}M_i
\end{eqnarray}
where $M_i=A_i\circ B_i$. For $i\neq 0$, the blocks, $M_i$ are all positive definite, since the corresponding matrices $A_i$ are positive definite. When we consider the new partially transposed matrix for (\ref{mix}), we have
\begin{eqnarray}
\rho'^{T_B} = (1/2)(\rho^{T_B}+|\Omega\rangle\langle\Omega|^{T_B}).\label{eqt}
\end{eqnarray}
We will now consider that $|\omega|$ is sufficiently small that we can neglect terms $O(|\omega^2|)$; we will discuss more on this point further in this Appendix. Thus, in this approximation we can say that effectively we have  
$|\Omega\rangle\langle\Omega|^{T_B}\approx (\sqrt{1-|\omega|^2} )(|00\rangle\langle 00|+\omega |\phi_{20}\rangle\langle 00|+\omega |00\rangle\langle\phi_{20}|)$.
We can rewrite Eq. (\ref{eqt}) as 
\begin{eqnarray}
\rho'^{T_B} = (1/2)\left[\left(\bigoplus_{i=-\infty}^{\infty}M'_i\right)+P\right],
\end{eqnarray}
where $M'_i=M_i$, for $i\neq 0$ and $M'_0=M_0+\sqrt{1-|\omega|^2}|00\rangle\langle 00|$, while $P=\omega(|\phi_{20}\rangle\langle 00|+|00\rangle\langle\phi_{20}|)^{T_B}$ represents a perturbation with magnitude totally dependent on the value of $|\omega|$. 

It is straightforward that for values $\sqrt{1-|\omega|^2} >1/2$, which means $|\omega|<\sqrt{3/4}$, the block $M'_0$ becomes positive-definite. We must know how small $|\omega|$ must be so that $\rho'^{T_B}$ remains positive. 

\textbf{Remark:} \textit{Since the eigenvalues of a matrix are continuous functions of its elements, we could already stop the demonstration at this point: slight variations of $\omega$ would not affect the eigenvalues of the positive blocks $M'_i$ significantly and consequently its positivity. However, the constructive demonstration given here has the advantage of giving an estimate on the order of magnitude of the value $\omega$, which is relevant experimentally}.

We need the following theorem (Theorem $6.1.1$ from \cite{horn}), known as the Gerschgorin Disc Theorem:
\begin{theorem}
Given a $n\times n$ square matrix $M=[m_{ij}]$, let
\begin{eqnarray}
R_i(M)\equiv \sum_{j=0,j\neq i}^{n-1} |m_{ij}|, \ \ i= 0, 1, \ldots, n-1, 
\end{eqnarray}
denote the deleted absolute row sums of $M$. Then all the eigenvalues of $M$ are located in the union of $n$ discs
\begin{eqnarray}
\bigcup_{i=0}^{n-1}\{z\in\mathcal{C}:|z-a_{ii}|\leq R_i(M)\}\equiv G(M).
\end{eqnarray}
Furthermore, if a union of $k$ of these $n$ discs form a connected region that is disjoint from all the remaining $n-k$ discs, then there are precisely $k$ eigenvalues of $M$ in this region. 
\end{theorem}
The region $G(M)$ is called the Gerschgorin region of $M$ and the individual discs in $G(M)$ are called Gerschgorin discs; the boundaries of these discs are called the Gerschgorin circles. A similar result holds for the collum-sums (Corollary $6.1.3$ from \cite{horn}), but since we are dealing with Hermitean matrices it will not affect the results. We need also the following refinement (Corollary $6.1.6$ from \cite{horn}):
\begin{corollary}
Let $D=diag\{d_0,d_1,\ldots\}$, with $d_i$ positive real numbers for all $i=0,1,\ldots$. Then all eigenvalues of $M$ lie in the region 
\begin{eqnarray*}
\bigcup_{i=0}^{n-1}\{z\in\mathcal{C}:|z-a_{ii}|\leq \frac{1}{d_i}\sum_{j=0, j\neq i}^{n-1}d_j|m_{ij}|\}\equiv G(D^{-1}MD).
\end{eqnarray*}
Moreover, the spectrum of $M$ is precisely the set $\bigcap_DG(D^{-1}MD)$.
\end{corollary}
Since the blocks $M'_i$ are all positive-definite, there exists a set of $d_i$s above which will bring all Gerschgorin disks to the positive segment of the real line. Also, there is a continuous of such $d_i$'s and we conclude then that a slight change in the row sums $R_i$ - which are constituted by the off-diagonal elements of the matrix - will not change the eigenvalues of a matrix, since this would correspond to a negligible deformation of the corresponding Gerschgorin region. Let us see how this applies to example (\ref{mix}).

The perturbation matrix $P$ affects only the first row and collum of each $M'_0$, $M'_{\pm 1}$ and $M'_{\pm 2}$. For the first lines of these blocks we have
\begin{eqnarray}
& & |z_0-\sqrt{1-{\omega}^2}|\leq 2\omega+R_0 \\
& & |z_1-\frac{p_1}{2}|\leq \omega+R_1 \\
& & |z_2-\frac{p_2}{4}|\leq \omega +R_2
\end{eqnarray}
By putting the sole value $\omega$, instead of the actual terms $\omega\cos^2(\theta)$, $\omega\sin^2(\theta)$ and $\omega\sin(\theta)\cos^2(\theta)$, we are doing an overstimation of the perturbation. By the equations above, if $|\omega|<< p_2/4 < p_1/2 < \sqrt{1-|{\omega}|^2}$, we will have that $\rho'^{T_B}$ will remain positive, since the associated Gerschgorin regions will be effectivelly unaffected. The value $p_2/4=1/16$ has a order of magnitude of $10^{-2}$; thus we will impose for $|\omega|$ a conservative upeer bound of $10^{-3}$. We can now justify the neglecting of terms $O(\omega^2)$: their rate of decrease is much faster than the rate of decrease of diagonal elements; also, we have not considered terms $\sin^k(\theta)\cos^{1-k}(\theta)$, which would make this rate even faster.


\begin{thebibliography}{52}
\bibitem{nota1} A Hankel matrix is a matrix whose coefficients depend on the sum of rows and columms, i.e., $h_{ij}=h_{i+j}$, having thus constant elements along skew-diagonals.
\bibitem{nota2}The Hadamard product of a positive-definite matrix with the rank-$1$ matrices of the form $D_j$, $E_j$ and $F_j$ does not affect the leading principal minors' positivity: multipliyng rows of a square matrix by a positive constant do not change the sign of the determinant of this matrix. This implies that positive-definiteness is preserved.
\bibitem{nota3} For $\theta =\pi/4$ there is not a contradiction and state (\ref{gclass}) would be permutationally symmetric in the sense of \cite{guhne}. In this reference, there is a method to construct entanglement witnesses to detect entanglement of permutationally symmetric states. We will let the separability of this special case as an open (and interesting) question.
\bibitem{nota4} This comes from two reasons. First, the underlying Hankel structure of the partially transposed matrix imposes a numerical ill-conditioning (see \cite{fasino}). The second and more severe reason is that there are no general results about the eigendecomposition of a Hadamard product of two matrices. So, even if one finds a good Hankel matrix structure, the various Hadamard products can spoil the whole process. 
Another related issue is that bipartite bound entanglement becomes rare with increasing Hilbert space dimension\cite{hlewenstein2}: even if one finds a bound entangled state it will probably be not very robust against small fluctuations.
\end{thebibliography}
\end{document}